\documentclass[iop,numberedappendix,twocolumn]{emulateapj}

\usepackage{array} 
\usepackage{amsfonts}
\usepackage{amssymb} 
\usepackage{amsmath}
\usepackage{graphicx}
\usepackage{enumerate}
\usepackage{verbatim}
\usepackage{color}
\usepackage{mathrsfs}
\usepackage{comment}
\usepackage{multirow} 
\usepackage{diagbox}
\bibpunct{(}{)}{;}{a}{}{,} 
\usepackage{url}
\usepackage{hyperref}
\usepackage{bm}
\usepackage{enumerate}
\usepackage{mathrsfs}
\usepackage{comment} 

\newcommand{\oum}{'Oumuamua} 
\newcommand{\mathbfss}[1]{\bm{\mathsf{#1}}}
\def\aj{{AJ}}                   
\def\apj{{ApJ}}                 
\def\apjl{{ApJ}}                
\def\apjs{{ApJS}}               
\def\aap{{A\&A}}                
\def\apss{{Ap\&SS}}          
\def\mnras{{MNRAS}}             
\def\nat{{Nature}}              

\def\psj{{PSJ}}

\shorttitle{On the non-gravitational acceleration of \oum{}}
\shortauthors{F.~Spada}

\begin{document}

\title{Revisiting the trajectory of the interstellar object 'Oumuamua: \\ preference for a radially directed non-gravitational acceleration?}
\author{Federico Spada}
\email{federico.spada@gmail.com}

\begin{abstract}
I present a re-analysis of the available observational constraints on the trajectory of \oum{}, the first confirmed interstellar object discovered in the solar system.
\oum{} passed through the inner solar system on a hyperbolic (i.e., unbound) trajectory. Its discovery occurred after perihelion passage, and near the time of its closest approach to Earth. 
After being observable for approximately four months, the object became too faint and was lost at a heliocentric distance of around $3$ au.

Intriguingly, analysis of the trajectory of  \oum{} revealed that a dynamical model including only gravitational accelerations does not provide a satisfactory fit of the data, and a non-gravitational term must be included.
The detected non-gravitational acceleration is compatible with either solar radiation pressure or recoil due to outgassing. 
It has, however, proved challenging to reconcile either interpretation with the existing quantitative models of such effects without postulating unusual physical properties for \oum{} (such as extremely low density and/or unusual geometry, non-standard chemistry).

My analysis independently confirms the detection of the non-gravitational acceleration.
After comparing several possible parametrizations for this effects, I find a strong preference for a radially directed non-gravitational acceleration, pointing away from the Sun, and a moderate preference for a power-law scaling with the heliocentric distance, with an exponent between $1$ and $2$.
These results provide valuable constraints on the physical mechanism behind the effect; a conclusive identification, however, is probably not possible on the basis of dynamical arguments alone.
\end{abstract}


\section{Introduction}
\label{introduction}



In recent years, after having been theoretically predicted for decades \citep[e.g.,][]{McGlynn_Chapman:1989}, the first interstellar objects passing through the solar system on unbound trajectories have finally been detected.
The first confirmed object, 1I/\oum{} \citep{Williams_ea:2017}, was discovered in late 2017 after its perihelion passage and near its closest approach to Earth. 
It appeared as a point-like source at all times, with no signs of cometary activity \citep{Jewitt_ea:2017,Meech_ea:2017,Trilling_ea:2018}.
In sharp contrast, the second interstellar interloper, 2I/Borisov, discovered in 2019 well in advance of its perihelion passage, displayed a bright coma of ejected dust \citep{Guzik_ea:2020}.
In addition, the first candidate meteor of probable interstellar origin has also been recently confirmed \citep{Siraj_Loeb:2022a}.

These discoveries open unprecedented possibilities to study and characterise (potentially, even in situ: \citealt{Seligman_Laughlin:2018, Hibberd_ea:2022, Siraj_ea:2022}) objects from other stellar systems, providing invaluable constraints on the nature and the formation processes of the constituents of planetary and stellar systems, as well as the interstellar medium.
For recent reviews on this nascent field, see \citet{Jewitt_Seligman:2022, MoroMartin:2022, Siraj_Loeb:2022b}.

Soon after its discovery, the analysis of the trajectory of \oum{} revealed the presence of a non-gravitational component in its acceleration \citep{Micheli_ea:2018}.
Notably, the physical interpretation of this phenomenon has remained challenging since its detection.
Indeed, the two leading explanations that have been proposed so far, namely, solar radiation pressure and recoil from comet-like outgassing, both run into difficulties when a quantitative description is attempted.

For the solar radiation pressure interpretation to be viable, a very low density of the object, several orders of magnitude less than that of water, must be invoked \citep{Micheli_ea:2018}. 
This could imply that \oum{} has a highly porous, fractal-like structure \citep{MoroMartin:2019}, or a very unusual geometry (similar to a lightsail: \citealt{Bialy_Loeb:2018}).
The absence of observed cometary activity, on the other hand, is an obvious problem for the outgassing explanation.
Again, physical properties unlike those of solar system comets must be postulated (\citealt{Seligman_Laughlin:2020}; \citealt{Seligman_ea:2021} see also \citealt{Jewitt_Seligman:2022}, and references therein),
although it should be noted that non-gravitational accelerations in solar system bodies that do not show sign of cometary activity have been recently reported by \citet{Seligman_ea:2023} and \citet{Farnocchia_ea:2023}.
In summary, the physical interpretation of the non-gravitational acceleration is still a debated, currently unsettled issue (most recently, see, for instance, \citealt{Bergner_Seligman:2023}, and \citealt{Hoang_Loeb:2023}).

Given the implications for the potential discovery of novel phenomena and for the overall understanding of a new class of objects, it is important to re-examine critically the evidence in support of the detection of a non-gravitational acceleration in \oum{}.
In this paper, I perform such a re-analysis, also testing alternative possibilities \citep[see, e.g.,][]{Katz:2019}.
Another goal of this paper is to study different parametrizations of the non-gravitational acceleration, and to compare their relative performance in fitting the data.
In particular, modelling features such as, for instance, the direction along which the non-gravitational acceleration acts with respect to the instantaneous position on the trajectory, or its functional dependence on the heliocentric distance, can provide insight critical to the identification of the physical mechanism behind this effect.

The paper is organised as follows: in Section \ref{data} I briefly discuss the astrometric observations on the trajectory of \oum{} available from the literature; in Section \ref{methods} I describe the dynamical model and the fitting procedure; in Section \ref{results} I present the results of the analysis, and I discuss their implications in Section \ref{discussion}. 

\section{Observational constraints}
\label{data}

The observational dataset constraining the trajectory of \oum{} consists of astrometric positions covering the time interval from its discovery to when the object faded below detectability (approximately 2017 October 14 to 2018 January 2).
The measurements were obtained from both ground-based facilities and from the Hubble Space Telescope (HST).

The astrometric positions of \oum{}, expressed as right ascension and declination reduced to the J2000.0 reference frame, are available from the Minor Planet Center (MPC) database\footnote{https://minorplanetcenter.net/tmp/1I.txt}.
In the analysis presented here, I have used the version of the data which was published as supplementary material in \citet{Micheli_ea:2018}. 
These comprise 178 ground-based measurements from 27 observing stations (after discarding 7 pairs which had been deemed unreliable by their respective observers), and 30 HST measurements, or 416 scalar measurements in total.

The geocentric position of the observer is also needed to obtain accurate line-of-sight information.   
For the HST observations, the geocentric location of the spacecraft at the time of the observations was also provided by \citet{Micheli_ea:2018}.
For the ground-based observations, I have used the information on the geocentric positions of the observing stations provided by the MPC database\footnote{\url{https://minorplanetcenter.net/iau/lists/ObsCodesF.html}}.
For each measurement I have adopted the uncertainty recommended by \citet{Micheli_ea:2018}.

\section{Methods}
\label{methods}

\subsection{Orbit determination procedure}

To determine the trajectory of \oum{} I have implemented from first principles an orbit determination procedure, consisting of two main steps \citep[see, e.g.,][]{Es76}: (i) preliminary orbit determination; (ii) differential correction of the orbit.
The procedure is sufficiently general to handle both bound and unbound heliocentric orbits, and could be readily modified to include additional effects in the force model, or to study the trajectory of different objects.

\subsubsection{Preliminary orbit determination}

The preliminary orbit determination is based on the assumption of two-body (Keplerian) motion; in other words, all perturbations are neglected and only the gravitational attraction from the primary body (in this case, the Sun) is retained.
The relation between the initial position and velocity vectors, ${\bm r}_0$ and ${\bm v}_0$, at an initial instant (or ``epoch") $t_0$, and the position and velocity, ${\bm r}$ and ${\bm v}$, at any other epoch $t$ is therefore analytical.

A Keplerian orbit is completely specified by six scalar quantities, the orbital elements.
For instance, the components of the initial state vector ${\bm y}_0 = ({\bm r}_0, {\bm v}_0)$ are a suitable choice of orbital elements.
Six scalar measurements, such as three pairs of right ascension and declination measurements at distinct epochs, are thus required, at a minimum, to determine an orbit. 

The preliminary orbit determination problem, as formulated above, is one of the fundamental problems of classical celestial mechanics and astrodynamics. 
The literature on the topic is extensive, and several methods of solution exist \citep[see, e.g.,][and references therein]{Es76, Ba, MG12, BMW}. 
In my calculations, I have used the approach described in chapter 2 of \citet{MG12}.
The initial determination of ${\bm r}_0$ and ${\bm v}_0$ is based on three observations at three different epochs selected from the available data.
Upon convergence, the algorithm yields the values of $\bm r$ and $\bm v$ at the central epoch. 
From this I obtain ${\bm r}_0$ and ${\bm v}_0$ by standard orbit propagation from the middle epoch to $t_0$, assuming a purely Keplerian orbit, consistently with the assumptions of the initial orbit determination itself.
For the Keplerian propagation I have implemented the universal variable formulation of \citet{BMW}.
The values of ${\bm r}_0$ and ${\bm v}_0$ obtained at this step are used as starting values for the following differential correction step. 
As a basic sanity check, I have verified that different choices of the three epochs used in the preliminary orbit determination do not affect the final results.

\subsubsection{Differential correction of the orbit}

In the second step, the preliminary orbit is improved iteratively.
The differential correction of the orbit uses all the available observational constraints. 
At this stage, the trajectory calculation also includes all the perturbations deemed relevant for the problem at hand, which thus requires resorting to numerical integration (see Section \ref{traj} below).

The differential correction technique is performed as follows (for full, didactic accounts, see \citealt{Es76}, \citealt{MG12}).
First, a vector of quantities which completely determines the numerically integrated trajectory is identified; for instance, ${\bm x} = ({\bm r}_0, {\bm v}_0, {\bm \psi})$, i.e., the initial state vector $({\bm r}_0, {\bm v}_0)$, extended to include any additional dynamical parameters $\bm \psi$ appearing in the equations of motion, whose values are also to be estimated.
For each particular instance of ${\bm x}$, the measurements at the available epochs can be predicted, and the predictions compared with their actually observed counterparts, forming the vector of observed--minus--computed measurements.
This vector of residuals, $\bm \nu$, thus encodes the information on the goodness of fit of the observations that is achieved by the particular choice of $\bm x$ for which it was calculated.
Since the dependence of $\bm \nu$ on $\bm x$ is nonlinear and overdetermined (i.e., there are usually more observations than parameters to be determined), it is in general not possible to find a closed--form solution for the $\bm x$ that drives the residuals to zero.
Instead, the correction of $\bm x$ is performed iteratively, with the objective to minimise a cost function defined in terms of $\bm \nu$ and of the uncertainties on the measurements.
In this sense, the differential correction of orbits is essentially an optimisation problem.

Adopting a cost function defined as: 
\begin{equation}
\label{cost}
Q = {\bm \nu}^T \mathbfss{W} {\bm \nu}, 
\end{equation}
where $\mathbfss{W}$ is a matrix whose diagonal elements are $1/\sigma_i^2$, with $\sigma_i$ being the uncertainties of the measurements, 
leads to the classical normal equations of the weighted least-squares method \citep[see, e.g.,][]{Farnocchia_ea:2015}: 
\begin{equation}
\label{eqnormal}
{\Delta \bm x} =  -(\mathbfss{B}^T\mathbfss{WB})^{-1} \mathbfss{B}^T\mathbfss{W}\, {\bm \nu}, 
\end{equation}
where ${\Delta \bm x}$ is the differential correction to be applied to $\bm x$, and 
the design matrix $\mathbfss{B} = \left( \frac{\partial {\bm \nu}}{\partial {\bm x}} \right)$
and the residuals $\bm \nu$ are both calculated using the current iteration of $\bm x$.

For details on the practical calculation of the residuals and of the design matrix implemented in my code, as well as a concise derivation of equation \eqref{eqnormal}, see Appendix \ref{app:residuals}.

\begin{figure*}
\begin{center}
\includegraphics[width=\textwidth,clip]{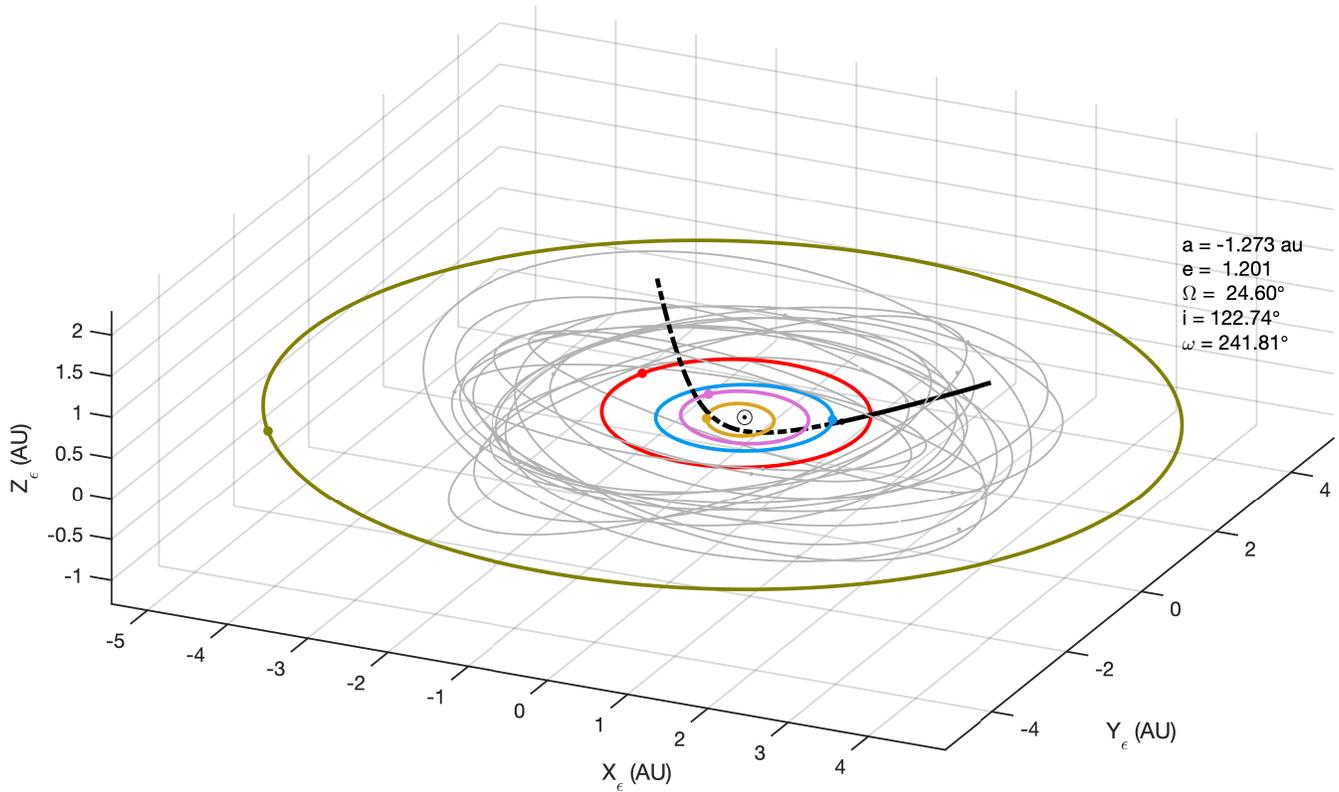}
\caption{Heliocentric ecliptic view of the trajectory of \oum{} through the inner solar system and beyond, plotted in black (solid line: from discovery to last observation; dash-dotted: pre-discovery, reconstructed). The planets from Mercury to Jupiter are shown in colours, while the sixteen main belt asteroids included in the model are plotted in grey; the Sun is represented with its symbol. The filled circles on the orbits show the positions of the respective objects at the time of discovery of \oum{} (2017 October 17); some of the osculating orbital elements at the same time are also displayed.}
\label{fig:trajectory}
\end{center}
\end{figure*}

\subsection{Trajectory model}
\label{traj}

The trajectory of \oum{} is modelled by direct numerical integration of the equations of motion:
\begin{flalign}
\label{EoM}
\begin{split}
{\dot {\bm r}} &= {\bm v}; \\ 
{\dot{\bm v}} &= {\bm a}(t,{\bm r},{\bm v}, {\bm \psi}),
\end{split}
\end{flalign}
where $\bm a$ denotes the acceleration of \oum{} in the heliocentric frame, and $\bm \psi$ represents additional dynamical parameters, such as those entering the definition of the non-gravitational forces.

The dynamical model includes the following purely gravitational effects: the Newtonian attraction from the Sun, which is the main term, the perturbations due to the eight major planets, Pluto, the Moon, and the sixteen most massive asteroids, as well as the relativistic correction term for the Sun. 

The gravitational perturbation from the bodies other than the Sun give rise to terms of the type \citep[e.g.,][]{Es76, MG12}:
\begin{equation}
{\bm a}_i = \mu_i \frac{{\bm r}_i - {\bm r}}{||{\bm r}_i - {\bm r}||^3} - \frac{{\bm r}_i}{||{\bm r}_i||^3},
\end{equation}
where $i$ is a running index identifying the perturber, $\mu_i = GM_i$ is its gravitational parameter (equal to its mass multiplied by the gravitational constant $G$), and ${\bm r}_i$ is the heliocentric position of the perturber body.
During the integration, the positions of all the solar system objects were obtained from the DE440 JPL ephemerides \citep{Park_ea:2021}.
The asteroids included are those listed in Table 2 of \citet{Farnocchia_ea:2015}, but with gravitational parameters updated to be consistent with the DE440 ephemerides.

The relativistic correction term is based on the post-Newtonian approximation \citep[see, e.g., equation 3.146 of][]{MG12}:
\begin{align}
\label{GR}
{\bm a}_{\rm rel} = \frac{\mu_\odot}{r^2}\left[ \left( 4\frac{\mu_0}{c^2 r} - \frac{v^2}{c^2} \right) {\bm e}_r + 4\frac{v^2}{c^2} ({\bm e}_r \cdot {\bm e}_v) {\bm e}_v \right],
\end{align}
where $\mu_\odot \equiv G M_\odot$ is the gravitational parameter of the Sun, and ${\bm e}_r = {\bm r}/r$, ${\bm e}_v = {\bm v}/v$ are unit vectors in the radial and tangential direction, respectively.

When non-gravitational terms are included, I used the general form \citep{Marsden_ea:1973}:
\begin{align}
\label{nongrav}
{\bm a}_{\rm NG} = g(r) \, ( A_1 {\bm e}_1 +  A_2 {\bm e}_2 + A_3 {\bm e}_3),
\end{align}
where 
\begin{align}
\label{gfun}
g(r) = (1\, {\rm au}/r)^k,
\end{align}
so that $A_1$, $A_2$, and $A_3$ represents the magnitudes at $1$ au of the components of the acceleration along the orthonormal vectors $\{ {\bm e}_1  {\bm e}_2  {\bm e}_3 \}$.
In the following, I have considered purely radial and purely along-track accelerations, i.e., acting along the direction of ${\bm e}_r$ and ${\bm e}_v$, respectively, as well as more general decompositions in the radial--transverse--normal (or RTN) basis:
\begin{flalign}
{\bm e}_1 = {\bm e}_{\rm R} \equiv {\bm e}_r; \ \ \ \ 
{\bm e}_3 = {\bm e}_{\rm N} \equiv {\bm e}_r \times {\bm e}_v; \ \ \ \ 
{\bm e}_2 = {\bm e}_{\rm T} \equiv {\bm e}_3 \times {\bm e}_1,
\end{flalign}
and the along-track--cross-track--normal basis (ACN):
\begin{flalign}
{\bm e}_1 = {\bm e}_{\rm A} \equiv {\bm e}_v; \ \ \ \ 
{\bm e}_3 = {\bm e}_{\rm N}; \ \ \ \ 
{\bm e}_2 ={\bm e}_{\rm C} \equiv {\bm e}_3 \times {\bm e}_1.
\end{flalign}

In summary, the full acceleration, including the solar main term and the perturbation terms, written in the heliocentric frame, reads \citep{Es76,Tr23}:
\begin{align}
\label{accel}
{\bm a} = - \frac{\mu_\odot}{||{\bm r}||^3} {\bm r} + 
\sum_i \mu_i \left[ \frac{{\bm r}_i - {\bm r}}{||{\bm r}_i - {\bm r}||^3} - \frac{{\bm r}_i}{||{\bm r}_i||^3} \right] 
+ {\bm a}_{\rm rel} + {\bm a}_{\rm NG},
\end{align}
where the sum in the second term is extended to the major planets, Pluto, the Moon, and the selected asteroids; their gravitational parameters $\mu_i$ are assigned consistently with the DE440 ephemerides \citep{Park_ea:2021}.

\begin{figure*}
\begin{center}
\includegraphics[width=0.49\textwidth,clip]{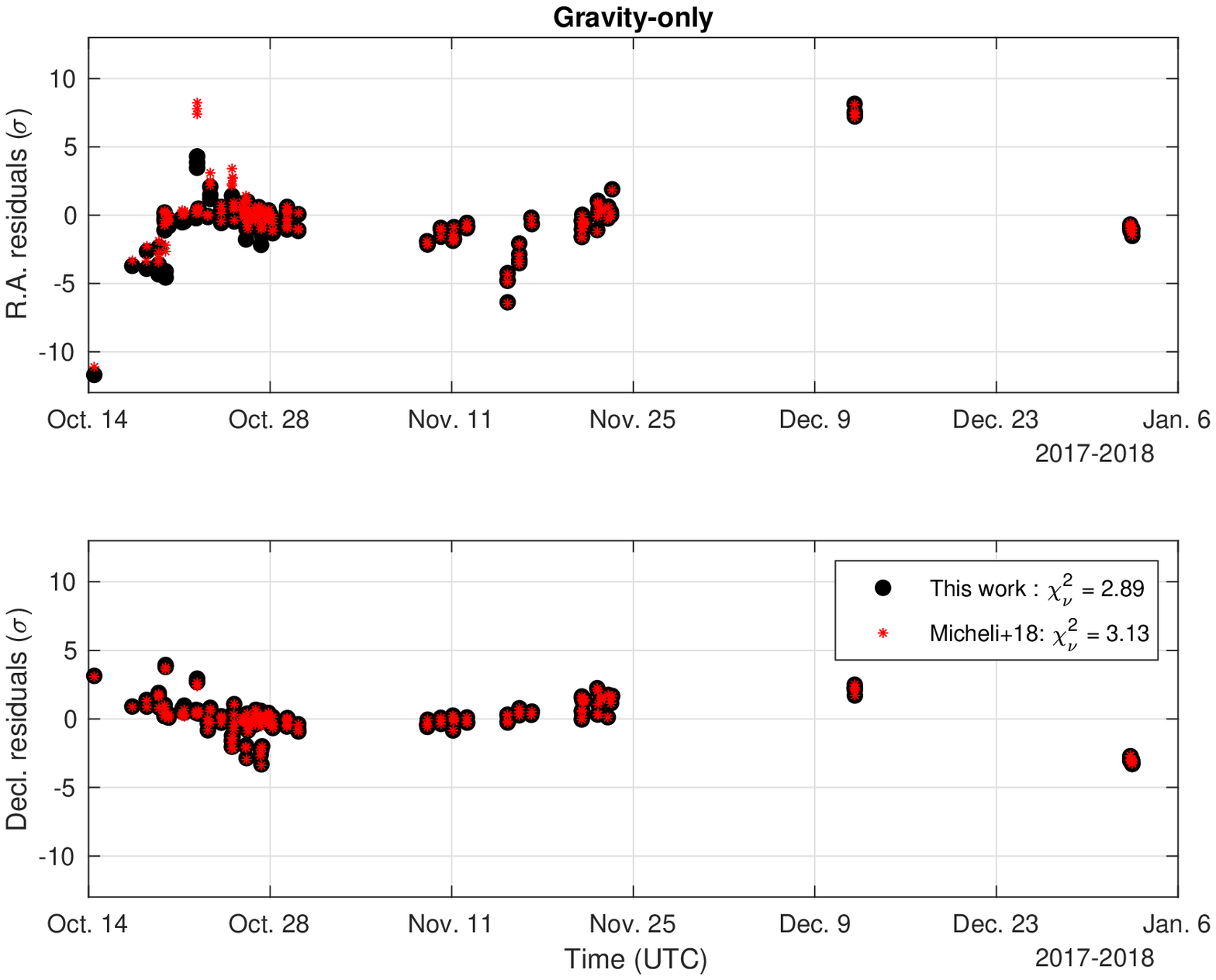}
\includegraphics[width=0.49\textwidth,clip]{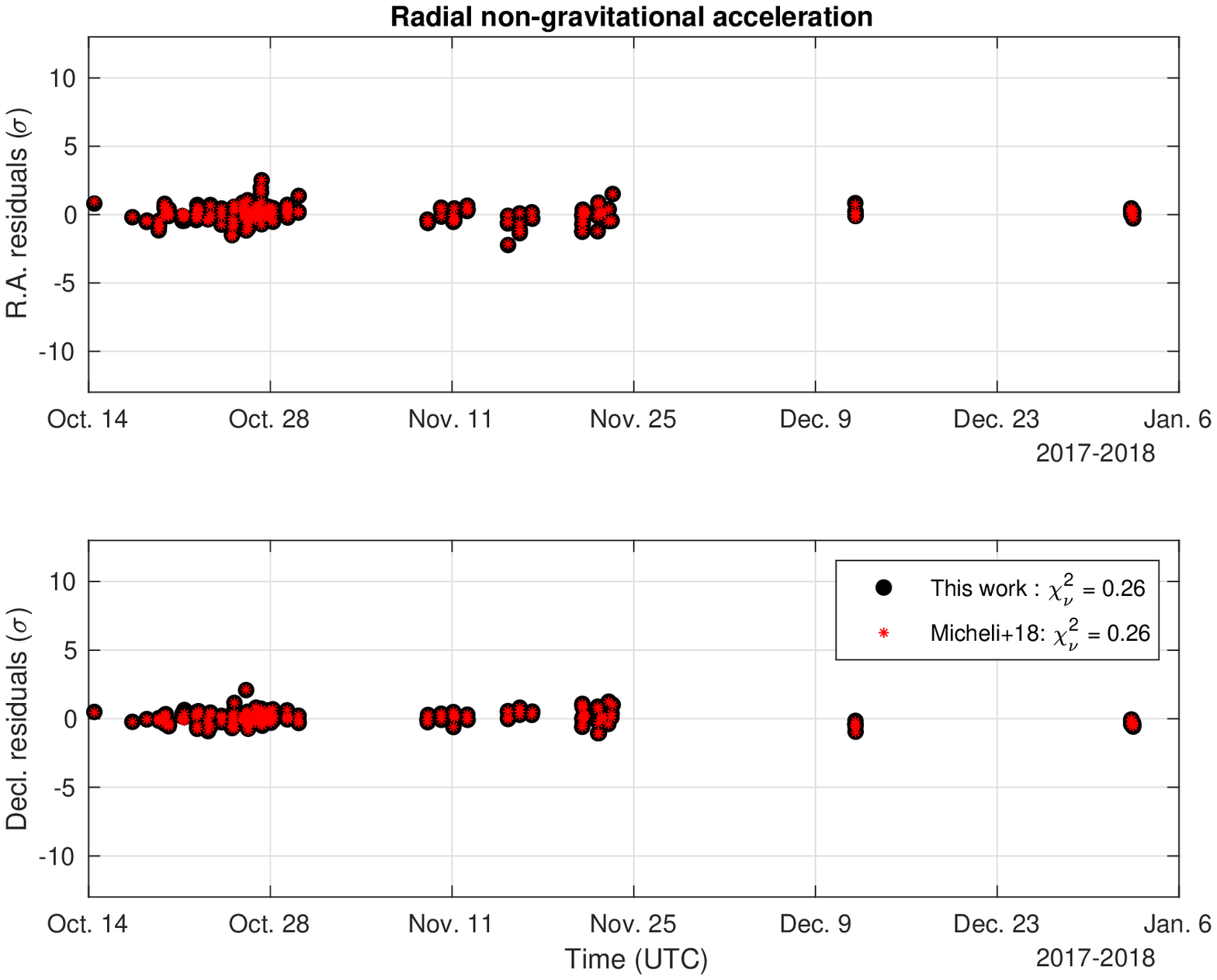}
\caption{Normalised residuals in right ascension and declination (top and bottom rows, respectively), for the gravity-only model (left panels), and a model including a radial non-gravitational acceleration (right panels). 
}
\label{fig:residuals}
\end{center}
\end{figure*}

\subsection{Numerical implementation}

The numerical analysis was entirely performed within the \mbox{MATLAB} computational environment.
To integrate the equations of motion \eqref{EoM}, I used the \texttt{ode113} solver, which implements a variable-step, variable-order Adams-Bashforth-Moulton predictor-corrector (PECE) solver (\citealt{Shampine_Reichelt:1997}; see also \citealt{SG75}).
For basic spherical astronomy computations, such as time systems and reference frames conversions, as well as to obtain the positions of solar system objects from the JPL DE440 ephemerides, I relied on the SPICE toolkit \citep{Acton:1996,Acton_ea:2018} accessed through its MATLAB interface, Mice\footnote{\url{https://naif.jpl.nasa.gov/pub/naif/toolkit_docs/MATLAB/}}.
The remaining standard astrodynamics calculations, such as Keplerian propagation, preliminary orbit determination, and orbit differential correction, were implemented from first principles.

Figure \ref{fig:trajectory} illustrates the reconstructed best-fitting trajectory of \oum{} through the inner solar system.
The Sun, the major planets up to Jupiter, and the selected asteroids included in the acceleration, equation \eqref{accel}, are shown with their orbits, with a circle marking their position at the time of discovery of \oum{}.

\section{Results}
\label{results}

\subsection{Revisiting the detection of non-gravitational acceleration}

As was shown by \citet{Micheli_ea:2018}, fitting the \oum{} astrometric observations to a gravity-only model, with ${\bm a}_{\rm NG}=0$ in equation \eqref{accel}, results in large right ascension and declination residuals, i.e., five times, or more of their formal uncertainty. 
In contrast, a model including a radial non-gravitational acceleration, with ${\bm a}_{\rm NG} = A_1 \left(1\, \rm au/r \right)^2 {\bm e}_r$, produces a satisfactory fit at the cost of only one additional free parameter, $A_1$.
In this sense, contrasting the results of these two fits constitutes the most economical indirect detection of a non-gravitational acceleration acting on \oum{}.

Using the method described in Section \ref{methods}, I have reproduced the fits for the same two dynamical models, i.e., gravity-only and gravity plus radial non-gravitational acceleration.
Figure \ref{fig:residuals} shows the results of my fits compared to those of \citet{Micheli_ea:2018} in terms of the normalised residuals in right ascension and declination (cf. their Figure 2).
This comparison is also useful as a basic validation of my implementation.

For the gravity-only model (panels to the left of Figure \ref{fig:residuals}), some of the residuals in right ascension in my fit show moderate discrepancies with respect to those of \citet{Micheli_ea:2018}.
The quality of my fit as measured by the $\chi_\nu^2$ statistics, is, however, comparable to theirs\footnote{The value of the reduced $\chi_\nu^2$ quoted here for \citet{Micheli_ea:2018} is calculated directly from the residuals as published in their supplementary tables; for the gravity-only model, there is a small inconsistency with the number reported in Table 1 of their paper.}.
In particular, I recover the strong deviations, at the level of $5$--$10 \sigma$, in both right ascension and declination, occurring at the same epochs as theirs.
More importantly, the residuals are not distributed at random, but are suggestive of some trends that the purely gravitational model is unable to capture, for instance, in the observations around 2017 October 19--22, or 2017 November 15--23.
These trends and the overall pattern of distribution of the residuals in my fit closely match those found by \citet{Micheli_ea:2018}. 

For my fit with the non-gravitational model, the residuals are plotted in the panels to the right of Figure \ref{fig:residuals}; my estimate of the parameter $A_1$, obtained from the orbit differential correction procedure, is $A_1 = (4.90 \pm 0.15) \times 10^{-6}$ m s$^{-2}$.
For this fit, the residuals, the $\chi_\nu^2$ value, and the estimate of $A_1$ are all in excellent agreement with the results of \citet{Micheli_ea:2018}. 

In summary, my results support and independently confirm that a gravity-only model is insufficient to explain the observed trajectory of \oum{}, while a model including a radial non-gravitational acceleration gives a satisfactory fit of the data.

\subsection{Testing for a spurious detection}

The detection of a non-gravitational acceleration discussed in the previous Section has been called into question by \citet{Katz:2019}, arguing that, since the non-gravitational acceleration is a smoothly varying function of the heliocentric distance, and thus of the time, it is surprising that its inclusion in the fit results in such a marked reduction of the residuals relative to nearly simultaneous epochs.
To assess the validity of this criticism, I have performed the following numerical experiment, also suggested by \citet{Katz:2019}.
After adding random noise with an amplitude equal to three times their formal uncertainties to the high-residual observations, I have repeated the fit with both the gravity-only and the radial non-gravitational acceleration models.
The improvement of the fit due to the inclusion of the non-gravitational term, obtained when the real data are used, should not be observed when fitting the model to the altered test data.

\begin{figure}
\begin{center}
\includegraphics[width=0.49\textwidth,clip]{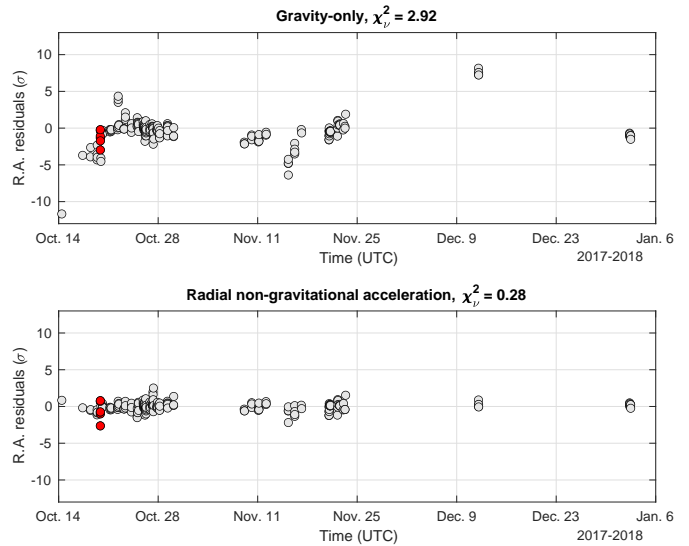}
\caption{Numerical experiment to test whether the improvement in the fit obtained with the non-gravitational dynamical model is due to spurious, non-physical reasons. 
Test data, to which white noise of amplitude $3\sigma$ was added, are shown in red; the remainder of the data are plotted in gray.}
\label{fig:katz}
\end{center}
\end{figure}

The results of this numerical experiment performed for the six observations obtained between 20:39 and 20:51 UTC of 2017 October 19 are illustrated in Figure \ref{fig:katz}.
For the test data, shown in red in the figure, the residuals of the fit including the non-gravitational term are approximately as large as for the gravity-only model. 
This is what would be expected if the improved fit of the non-gravitational model is not a spurious result, but is due to its ability to capture a genuine, physical effect, which the gravity-only model does not.
I have repeated the same numerical experiment for the high-residual observations corresponding to the nights of October 22, November 15, and December 12, and I have obtained similar results. 
I conclude that the concern reported above about the detection of a non-gravitational acceleration is unwarranted, and the existence of this additional effect is indeed supported by the data.

\subsection{Comparison of different parametrizations of the non-gravitational term}

Although a model including a purely radial non-gravitational acceleration suffices to obtain a satisfactory fit of the \oum{} trajectory with the least number of extra free parameters, it is interesting to explore other possible parametrizations of ${\bm a}_{\rm NG}$.
In the following I consider selected variations on the basic model discussed so far, namely, different values of the exponent $k$ in equation \eqref{gfun}, as well as different vector decompositions for the alignment of the vector ${\bm a}_{\rm NG}$.
The results of my fits are reported in Table \ref{tab:ngexpr}.

For the purely radial cases, all the values of power-law exponent $k$ in the range considered appear capable of fitting the data satisfactorily, although a slightly better performance is found for $k=1$ and $k=2$.
The quality of the fit deteriorates quickly for values of $k$ larger than $4$ (not reported in Table \ref{tab:ngexpr}).  
I have also considered two additional purely radial models, with a prescription of the function $g(r)$ based on empirical outgassing models of $\rm H_2O$ \citep{Marsden_ea:1973} and $\rm CO$ \citep{Yabushita:1996}.
In both cases, the empirical $g(r)$ functions have a radial profile similar to $(1\, {\rm au}/r)^2$ for $r < 1$ au, followed by steeper decline (significantly steeper in the $\rm H_2O$ case) at larger heliocentric distances.
Both the $A_1$ coefficients and the $\chi_\nu^2$ values roughly conform to the pattern followed by the power-law cases.
It is important to note that in all the purely radial models discussed so far $A_1 > 0$, in other words, the non-gravitational acceleration is directed away from the Sun.

In the case of the RTN decomposition, the optimised coefficients of the transverse and normal components of the acceleration are compatible with zero, while the values of $A_1$ and the $\chi_\nu^2$ are close to those of the purely radial case.

The purely along-track alignment of ${\bm a}_{\rm NG}$ is clearly disfavoured by the data.
The optimised values of the acceleration coefficient are very small or not significant, in other words, the best-fitting model is compatible with ${\bm a}_{\rm NG}=0$.
As a result, both the residuals and the $\chi_\nu^2$ values nearly coincide or are very close to those of the gravity-only model.

Finally, the ACN decomposition produces fits comparable to those obtained with the purely radial alignment, but with a tendency to non-significant, or very small, cross-track component.
This result can be understood as follows.
For most of the post-discovery trajectory of \oum{}, the radius vector and the velocity vector are roughly aligned (their mutual angle is always between 15 and 20 degrees); this fact can be also easily understood from Figure \ref{fig:trajectory}.
As a consequence, the along-track and cross-track directions are roughly aligned with, and orthogonal to, the radial direction, respectively.
The performance of the ACN models can thus be understood as a (sub-optimal) approximation of the purely radial models.

In conclusion, the purely radial non-gravitational acceleration is clearly favoured by the data. 
A moderate preference for a power-law dependence on the heliocentric distance, with a not too steep slope, is also visible, although other radial dependences cannot be ruled out.

\begin{table}
\caption{Best-fitting parameters and reduced $\chi_\nu^2$ statistics for different parametrizations of the non-gravitational acceleration ${\bm a}_{\rm NG}$ in equation \eqref{nongrav}. The coefficients $A_i$ and their uncertainties are in units of $10^{-6}$ m s$^{-2}$. Non-significant values (i.e., compatible with zero within $3\sigma$) are shown in parentheses. The number $k$ is the exponent appearing in equation \eqref{gfun}.
}
\begin{center}
\resizebox{\columnwidth}{!}{
\begin{tabular}{ccccc}
\hline
 $k$  &  $A_1$  &      $A_2$  &       $A_3$  &        $\chi_\nu^2$ \\
 \hline
 \hline
\multicolumn{5}{c}{Purely radial} \\
 \hline
 0  &    ${2.07\pm0.063}$  &   N/A   &  N/A   &  $0.29$ \\
 1  &    ${3.18\pm0.097}$  &   N/A   &  N/A   &  $0.25$ \\
 2  &    ${4.90\pm0.15}$    &   N/A   &  N/A   &  $0.26$ \\
 3  &    ${7.46\pm0.23}$    &   N/A   &  N/A   &  $0.31$ \\
 $\rm CO^\dagger$ & ${5.37\pm0.16}$ & N/A & N/A & $0.27$ \\
 $\rm H_2O^\ddagger$ & ${6.19\pm0.19}$ & N/A & N/A & $0.36$ \\
  \hline
 \hline
\multicolumn{5}{c}{RTN decomposition} \\
 \hline
 0  &    ${2.02\pm0.13}$  &   $(-0.090\pm0.11)$   &  $(-0.11\pm 0.11)$  &  $0.28$ \\
 1  &    ${3.19\pm0.29}$  &   $( 0.011\pm0.20)$   &  $( 0.012 \pm 0.20)$ &  $0.25$ \\
 2  &    ${5.00\pm0.56}$  &   $( 0.077\pm0.39)$   &  $( 0.12\pm 0.36)$ &  $0.25$ \\
 3  &    ${7.08\pm1.01}$  &   $(-0.35\pm0.74)$   &  $(-0.17\pm 0.66)$  &  $0.27$ \\
 \hline
 \hline
\multicolumn{5}{c}{Purely along-track} \\
 \hline
 0  &   $(-0.006\pm0.020)$        &   N/A    &   N/A &   $2.89$ \\
 1  &   $( 0.11\pm0.038)$           &  N/A    &   N/A &   $2.88$ \\
 2  &   ${0.42\pm0.070}$           &   N/A    &   N/A &   $2.81$ \\
 3  &   ${1.13\pm0.12}$             &   N/A    &   N/A &   $2.69$ \\
 \hline
 \hline
\multicolumn{5}{c}{ACN decomposition} \\
 \hline
 0  &    ${2.65\pm 0.22}$  &   ${-0.29\pm 0.092}$   &  ${0.45\pm 0.15}$  & $0.32$ \\
 1  &    ${4.87\pm 0.52}$  &   $(-0.21\pm 0.18)$     &  ${1.16\pm 0.31}$  & $0.26$ \\
 2  &    ${8.81\pm 1.09}$  &   $( 0.051\pm0.36)$    &  ${2.47\pm0.61}$   & $0.24$ \\
 3  &    ${14.09\pm2.14}$  &  $( 0.33\pm0.74)$      &  ${4.11\pm1.19}$   & $0.26$ \\
 \hline
 \hline
\end{tabular}
}
\end{center}
$^\dagger$: $g(r)$ according to equation (4.4) of \citet{Yabushita:1996};  \\
$^\ddagger$: $g(r)$ according to equation (5) of \citet{Marsden_ea:1973}.
\label{tab:ngexpr}
\end{table}%

\section{Discussion}
\label{discussion}

The fit of the dynamical models to the observed trajectory of \oum{} provides useful constraints on the nature of the physical mechanism responsible for the detected non-gravitational acceleration.
In general, the preference found for the models with radially directed non-gravitational acceleration, pointing away from the Sun, is suggestive of a physical process connected with the interaction with solar radiation, either directly (i.e., radiation pressure), or indirectly (outgassing induced by sublimation from the sunlit side of the object).

It should be noted that the effect of solar radiation pressure is naturally represented by a term of the form\footnote{At least, to the leading order, i.e., considering only the absorption of solar radiation; see, e.g., \citet{Veras_ea:2015} for a more complete description of the dynamical effects of the interaction with solar radiation.} $\propto \left(1\, \rm au/r \right)^2 {\bm e}_r$, while this is not necessarily true of the recoil due to outgassing.
In any case, it is not possible to rule definitively in favour of one of the two options on the basis of the results of Section \ref{results} alone.
Intriguingly, both interpretations run into difficulties when examined in the light of extant quantitative models.

Given the value of $A_1$, the radiation pressure interpretation requires postulating that \oum{} is several orders of magnitude less dense than water \citep{Micheli_ea:2018}.
This would imply a highly porous, fractal structure \citep{MoroMartin:2019}, which, although possible, has not been observed so far in the solar system in objects of comparable size.
Alternatively, \citet{Bialy_Loeb:2018} proposed that \oum{} has the morphology of a thin sheet (akin to a lightsail).

Measurements in solar system comets show that outgassing primarily occur due to sublimation from the hot ``dayside" of the nucleus; the recoil from such a process would thus be compatible with a primarily radial acceleration, directed away from the Sun. 
This explanation, however, faces difficulties in the identification of the sublimating volatile species, and due to the fact that a coma of dust, which normally accompanies outgassing in solar system comets, was not detected for \oum{} \citep[see][for more details and proposed solutions]{Seligman_Laughlin:2020, Jewitt_Seligman:2022}. 
It should also be noted that significant detections of non-gravitational accelerations in solar system bodies with no observed cometary activity have been recently reported by \citet{Seligman_ea:2023, Farnocchia_ea:2023}.

\begin{figure}
\begin{center}
\includegraphics[width=0.49\textwidth,clip]{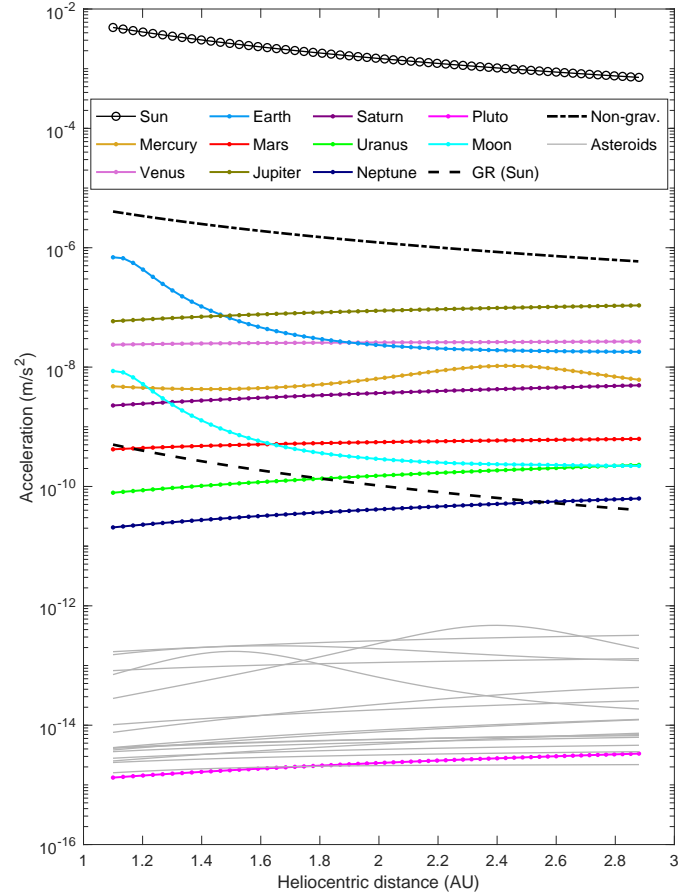}
\caption{Acceleration terms acting on \oum{} as functions of its heliocentric distance.
The label ``GR (Sun)" marks the general relativistic correction term due to the Sun (see equation \ref{GR}). 
The non-gravitational term (dash-dotted curve) is plotted according to the purely radial prescription, ${\bm a}_{\rm NG} = A_1 \left(1\, \rm au/r \right)^2 {\bm e}_r$, with $A_1 = (4.90 \pm 0.15) \times 10^{-6}$ m s$^{-2}$.}
\label{fig:accel}
\end{center}
\end{figure}

Finally, the magnitude of the non-gravitational effect can be put into context with respect to the other acceleration terms of gravitational origin with the help of Figure \ref{fig:accel}.
In the Figure, the non-gravitational term is plotted according to the purely radial, $r^{-2}$ prescription.
Clearly, $||{\bm a}_{\rm NG}||$ is the largest of the perturbative effects.
At $1$ au it amounts to approximately $0.1\%$ of the main term, the local solar acceleration.
Another way to look at this is the following: the non-gravitational acceleration is effectively equivalent to a reduction of the solar gravitational parameter by a factor of $0.99917$ \citep{Katz:2019}.
The standard two-body relation
\begin{equation}
v^2 = \mu_\odot \left( \frac{2}{r} - \frac{1}{a} \right),
\end{equation}
implies that:
\begin{equation}
\frac{\delta v}{v} = \frac{1}{2} \frac{\delta \mu_\odot}{\mu_\odot} \approx 4.15 \cdot 10^{-4}.
\end{equation}
In other words, $0.04\%$ of the velocity of \oum{} is due to the non-gravitational perturbation.

In comparison with solar system comets, the value of $A_1$ found for \oum{} is among the largest \citep{Micheli_ea:2018}, although \citet{Jewitt_Seligman:2022} noted that this may be mostly a result of its comparatively small size.

\section{Conclusions}
\label{conclusions}

In this paper I have re-analysed the existing astrometric observations of the trajectory of \oum{}, the first discovered interstellar object, on the basis of an independently implemented dynamical model.

My analysis confirms the detection of a non-gravitational component in the acceleration of \oum{}, as previously reported in the literature, after numerically testing, and rejecting, the possibility that it is a spurious artefact of the fitting procedure.

Among the parametrizations considered, I find that a non-gravitational acceleration acting radially and pointing away from the Sun fits the data best, while alternative alignments produce poorer fits and/or contain more free parameters.
For the model scaling with the inverse square of the heliocentric distance, i.e., ${\bm a}_{\rm NG} = A_1 (1\, {\rm au}/r)^2 {\bm e}_r$, my estimate of the magnitude of the acceleration at $1$ au is $A_1 = (4.90 \pm 0.15) \times 10^{-6}$ m s$^{-2}$.

My results are compatible with a non-gravitational acceleration arising from the interaction of the object with solar radiation, such as the effect of solar radiation pressure, or the recoil due to outgassing of sublimating material from the surface.

\section*{Acknowledgements}
This research has made use of NASA's Astrophysics Data System Bibliographic Services, and of the SPICE system software and the SPICE server provided by NAIF.


\appendix

\section{Practical details of the implementation of differential correction of orbits}
\label{app:residuals}

In this Appendix I describe some practical aspects of my implementation of the differential correction of orbits.
In particular, I give a full derivation of the fundamental relations used to calculate the residuals and their partial derivatives with respect to the state vector.

\subsection{Derivation of the normal equations for the differential correction of orbits}

The derivation of equation \eqref{eqnormal} is based on a linearisation procedure.
Calling $\bm z$ the vector of available measurements and ${\bm h}(\bm x)$ their computed counterparts expressed as a function of ${\bm x} = ({\bm r_0}, {\bm v}_0, {\bm \psi})^T$, the residuals can be expanded in a Taylor series around a reference value ${\bm x} = {\bm x}^*$ (e.g., its estimate at the current iteration):
\begin{align}
{\bm \nu}({\bm x}) = {\bm z} - {\bm h}({\bm x}) \approx {\bm z} - {\bm h}({\bm x}^*) - \left(\frac{\partial {\bm h}}{\partial {\bm x}}\right)_{{\bm x}^*} \Delta {\bm x} + \dots,
\end{align}
or, retaining only the linear term in $\Delta {\bm x}$:
\begin{equation}
{\bm \nu} = {\bm \nu}^* - \mathbfss{H} \, \Delta {\bm x},
\end{equation}
where $\mathbfss{H} = \left(\frac{\partial {\bm h}}{\partial {\bm x}}\right)_{{\bm x}^*} $.
Inserting this linearised expression in the definition of the cost function (see equation \ref{cost}):
\begin{equation*}
Q = \nu^T \mathbfss{W} \nu = ({\bm \nu}^* - \mathbfss{H} \, \Delta {\bm x})^T \mathbfss{W} ({\bm \nu}^* - \mathbfss{H} \, \Delta {\bm x}).
\end{equation*}
The ${\Delta \bm x}$ for which $Q$ is minimum satisfy the conditions:
\begin{flalign}
\frac{\partial Q}{\partial \Delta {\bm x}} &= 2\left( \mathbfss{H}^T \mathbfss{W} \mathbfss{H} \, {\Delta \bm x} - \mathbfss{H}^T\mathbfss{W} {\bm \nu}^* \right) = 0; \\
\frac{\partial^2 Q}{\partial \Delta {\bm x}^2} &=  2\, \mathbfss{H}^T \mathbfss{W} \mathbfss{H} \ \ {\rm positive \ definite}.
\end{flalign}
From the first condition, equation \eqref{eqnormal} immediately follows with ${\bm \nu} \equiv {\bm \nu}^*$ and $\mathbfss{B} \equiv \mathbfss{H}$.
The second condition is automatically satisfied if $\mathbfss{H}$ has full rank \citep{STB, MG12}.

\subsection{Calculation of the residuals}

In the angles-only case, the observations consist of pairs of right ascension and declination angles, $(\alpha_i, \delta_i)$, at each epoch $t_i$, with $i=1,\dots, N$ (i.e., $2N$ scalar observations in total).
From these, the vectors $A_i$, $D_i$:
\begin{equation}
{\bm A}_i = 
\begin{pmatrix} 
-\sin \alpha_i \\  
\phantom{-}\cos \alpha_i \\  
0 
\end{pmatrix}; \ \
{\bm D}_i = 
\begin{pmatrix} 
-\sin \delta_i \cos \alpha_i \\ 
-\sin \delta_i \sin \alpha_i \\ 
\cos \delta_i 
\end{pmatrix},
\end{equation}
as well as ${\bm L}_i^O$, the observed line of sight vector:
\begin{align}
{\bm L}_i^O = 
\begin{pmatrix}  
\cos \delta_i \cos \alpha_i \\ 
\cos \delta_i \sin \alpha_i \\ 
\sin \delta_i 
\end{pmatrix},
\end{align}
can be calculated at each epoch $t_i$.
By construction, the vectors ${\bm A}_i$ and ${\bm D}_i$ are orthogonal to ${\bm L}_i^O$.
Its computed counterpart, ${\bm L}_i^C$, is the unit vector in the direction of the slant range vector $\bm \rho_i$, i.e., ${\bm L}_i^C = {\bm \rho}_i / \rho_i$. 
The range vector is in turn derived from the numerically integrated trajectory, ${\bm r}(t)$, and the heliocentric observer position vector, $\bm R$:
\begin{align}
\label{range}
{\bm \rho}_i = {\bm r} \left(t_i - \frac{\rho_i}{c} \right) - {\bm R}_i(t_i).
\end{align}

Two subtle points arise in dealing with equation \eqref{range}.
First, note that the position along the trajectory is evaluated at the time $t_i - \frac{\rho_i}{c}$, which is the epoch time decremented by the light travel time $\rho_i/c$. 
Accounting for the finite speed of light is necessary for accurate orbital calculations; including this correction makes equation \eqref{range} implicit, thus requiring an iteration for its solution.
Secondly, the vector ${\bm R}_i$ is the heliocentric position of the observing station, which is the sum of the geocentric position vector of the observing station, ${\bm R}_{s,i}$, and the heliocentric position of the Earth at time $t_i$, ${\bm r}_\oplus(t_i)$:
\begin{align}
{\bm R}_i(t_i) = {\bm r}_\oplus(t_i) + {\bm R}_{s,i}(t_i).
\end{align}
When ${\bm R}_{s,i}$ is transformed from the standard Earth-fixed reference frame to the J2000 reference frame, the transformation should account for the non-spherical shape of the Earth and its orientation at time $t_i$ (i.e., the rotation with respect to the J2000 frame, as well as the effect of nutation and precession).

Finally, the residuals in right ascension and declination can be expressed in terms of ${\bm A}_i$, ${\bm D}_i$, and the line of sight vectors ${\bm L}_i^O$ and ${\bm L}_i^C$ as follows (note that the right ascension residuals contain the geometric factor $\cos \delta_i$):
\begin{align}
{\bm \Delta L}_i = ({\bm L}_i^O - {\bm L}_i^C) = (\Delta \alpha_i \cos \delta_i) {\bm A}_i + (\Delta \delta_i) {\bm D}_i ,
\end{align}
so that, using the orthogonality of ${\bm A}_i$ and ${\bm D}_i$, we obtain explicit formulae for the components of the vector of the residuals:
\begin{flalign}
\label{residuals}
\begin{split}
\nu_i &= (\cos \delta_i \Delta \alpha_i) = {\bm A}_i \cdot {\bm \Delta L}_i \ \ \ {\rm for} \ \ \  i=1,\dots, N;
\\ 
\nu_i &= (\Delta \delta_i) = {\bm D}_i \cdot {\bm \Delta L}_i \ \ \ {\rm for} \ \ \  i=N+1,\dots,2\, N. 
\end{split}
\end{flalign}

\subsection{Partial derivatives of the residuals with respect to the state vector}

Starting from the basic relation, valid for any epoch:
\begin{align}
\rho {\bm L} = {\bm \rho} = ({\bm r} - {\bm R}),
\end{align}
and denoting with $\xi$ a generic component of the state vector ${\bm y} = ({\bm r}, {\bm v})$, we can write, since $\bm R$ does not depend on $\xi$:
\begin{align}
\frac{\partial}{\partial \xi}(\rho {\bm L}) = \frac{\partial \rho}{\partial \xi} {\bm L} +  \rho \frac{\partial {\bm L}}{\partial \xi} = \frac{\partial {\bm r}}{\partial \xi}.
\end{align}
Exploiting the relation \citep[cf. equation 9.140 of][]{Es76}:
\begin{align}
\frac{\partial \rho}{\partial \xi} = {\bm L} \cdot \frac{\partial {\bm r}}{\partial \xi},
\end{align}
we obtain:
\begin{align}
\frac{\partial {\bm L}}{\partial \xi} =  - \frac{{\bm L}}{\rho} \left( {\bm L} \cdot \frac{\partial {\bm r}}{\partial \xi} \right) + \frac{1}{\rho} \frac{\partial {\bm r}}{\partial \xi}.
\end{align}
Using the fact that ${\bm A}$ and ${\bm D}$ are both orthogonal to ${\bm L}$, we obtain the very simple relations:
\begin{align}
\begin{split}
{\bm A} \cdot \frac{\partial {\bm L}}{\partial \xi} = \frac{{\bm A}}{\rho} \cdot \frac{\partial {\bm r}}{\partial \xi}; \\ 
{\bm D} \cdot \frac{\partial {\bm L}}{\partial \xi} = \frac{{\bm D}}{\rho} \cdot \frac{\partial {\bm r}}{\partial \xi}.
\end{split}
\end{align}
Recalling equations \eqref{residuals}, and taking into account that the partial derivatives of the observed quantities with respect to $\xi$ are zero, we arrive at the following expressions for the partials of the residuals with respect to a component of the state vector:
\begin{flalign}
\begin{split}
\frac{\partial \nu_i}{\partial \xi} &= \frac{{\bm A}_i}{\rho_i} \cdot \frac{\partial {\bm r}_i}{\partial \xi} \ \ \ {\rm for} \ \ \  i=1,\dots, N;
\\
\frac{\partial \nu_i}{\partial \xi} &= \frac{{\bm D}_i}{\rho_i} \cdot \frac{\partial {\bm r}_i}{\partial \xi} \ \ \ {\rm for} \ \ \  i=N+1,\dots, 2\, N,
\end{split}
\end{flalign}
or, in vector form:
\begin{flalign}
\label{partials_res}
\begin{split}
\frac{\partial \nu_i}{\partial {\bm y}(t_i)} &= \frac{1}{\rho_i} ( {\bm A_i} , \ {\bm 0}_{1\times3})^T, \ \ \ {\rm for} \ \ \  i=1,\dots, N; \\
\frac{\partial \nu_i}{\partial {\bm y}(t_i)} &= \frac{1}{\rho_i} ( {\bm D_i} , \ {\bm 0}_{1\times3})^T, \ \ \ {\rm for} \ \ \  i=N+1,\dots, 2\, N,
\end{split}
\end{flalign}
where, for instance, $( {\bm A_i} , \ {\bm 0}_{1\times3})^T$ is the $1\times 6$ row vector whose components are $(A_{i,x}, A_{i,y}, A_{i,z}, 0, 0, 0)$.

\subsection{State transition matrix and sensitivity matrix}

Equations \eqref{partials_res} can be rewritten in a more convenient form by introducing the state transition matrix $\bm \Phi$ and the sensitivity matrix $\mathbfss{S}$ \citep[see][]{MG12}.
The state transition matrix expresses the dependence of the state vector at time $t$ on the initial state vector at time $t_0$:
\begin{equation}
{\bm \Phi}(t,t_0) = \left( \frac{\partial {\bm y}(t)}{\partial {\bm y}(t_0)} \right) =
\begin{pmatrix}
\dfrac{\partial {\bm r}}{\partial {\bm r_0}} & \dfrac{\partial {\bm r}}{\partial {\bm v_0}} \\
\dfrac{\partial {\bm v}}{\partial {\bm r_0}} & \dfrac{\partial {\bm v}}{\partial {\bm v_0}} \\
\end{pmatrix}_{\rm 6 \times 6}.
\end{equation}
The sensitivity matrix expresses the dependence of the state vector on the force model parameters:
\begin{equation}
\mathbfss{S}(t) = \left( \frac{\partial {\bm y}(t)}{\partial {\bm \psi}} \right)_{\rm 6 \times n_p},
\end{equation}
where $n_p$ is the length of $\bm \psi$.
With these definitions, and using equations \eqref{partials_res}, the rows of the design matrix introduced in equation \eqref{eqnormal} can be written as:
\begin{equation}
\left( \frac{\partial \nu_i}{\partial {\bm x}} \right) = 
\left( \frac{\partial \nu_i}{\partial {\bm y}_0}, \  \frac{\partial \nu_i}{\partial {\bm \psi}} \right) =
\left(  \left( \frac{\partial \nu_i}{\partial {\bm y}} \right) \left( {\bm \Phi}(t_i,t_0) , \ \mathbfss{S}(t_i) \right)  \right),
\end{equation}
or:
\begin{flalign}
\label{design}
\begin{split}
\frac{\partial \nu_i}{\partial {\bm x}} &= \underbrace{( \frac{{\bm A_i}}{\rho_i} , \ {\bm 0}_{1\times3})^T}_{\rm 1 \times 6 \, {\rm row \, vector}}  \underbrace{\left( {\bm \Phi}(t_i,t_0) , \ \mathbfss{S}(t_i) \right)}_{6 \times (6+n_p) \, {\rm matrix}}  , \ \  {\rm for} \  i=1,\dots, N; \\
\frac{\partial \nu_i}{\partial {\bm x}} &= \underbrace{( \frac{{\bm D_i}}{\rho_i} , \ {\bm 0}_{1\times3})^T}_{\rm 1 \times 6 \, {\rm row \, vector}}   \underbrace{\left( {\bm \Phi}(t_i,t_0) , \ \mathbfss{S}(t_i) \right)}_{6 \times (6+n_p) \, {\rm matrix}} , \ \  {\rm for} \   i=N+1,\dots, 2\, N.
\end{split}
\end{flalign}
The practical importance of equations \eqref{design} stems from the fact that the state transition matrix and the sensitivity matrix can be readily obtained at any time, either analytically or numerically.
Equations \eqref{design} thus express the design matrix in terms of quantities that are all know at each step of the iteration in $\bm x$.

In my implementation, I have opted for a numerical calculation of $\bm \Phi$ and $\mathbfss{S}$, performed together with the trajectory integration described in Section \ref{traj}.
In practice, this means that the following matrix differential equations are integrated simultaneously along with equations \eqref{EoM}:
\begin{equation}
\label{phi_s}
\frac{d}{dt} \left( {\bm \Phi}, \ \mathbfss{S}  \right) = 
\begin{pmatrix}
\mathbfss{0}_{\rm 3 \times 3} & \mathbfss{1}_{\rm 3 \times 3} \\
\\
\dfrac{\partial {\bm a}}{\partial {\bm r}} & \dfrac{\partial {\bm a}}{\partial {\bm v}} \\
\end{pmatrix}
 \left( {\bm \Phi}, \ \mathbfss{S}  \right) +
\begin{pmatrix}
\mathbfss{0}_{\rm 3 \times 6} &  \mathbfss{0}_{\rm 3 \times n_p} \\
\\
\mathbfss{0}_{\rm 3 \times 6}  & \dfrac{\partial {\bm a}}{\partial {\bm \psi}} \\
\end{pmatrix},
\end{equation}
the initial conditions for equations \eqref{phi_s} being ${\bm \Phi}(t_0,t_0) = \mathbfss{1}$ and $\mathbfss{S}(t_0) = \mathbfss{0}$.
Of course, for the gravity-only model, $n_p=0$ and the sensitivity matrix does not appear in the calculations. 

For most of the acceleration terms included in equation \eqref{accel} analytical expressions for $\frac{\partial {\bm a}}{\partial {\bm r}}$ are available (see equations 7.75 and 7.77 of \citealt{MG12}). 
Also note that the general relativistic correction is the only term that would contribute to $\frac{\partial {\bm a}}{\partial {\bm v}}$; such small term can also be omitted in the calculation of $\bm \Phi$ without significantly affecting the convergence of the differential correction procedure.

When the non-gravitational acceleration term is included, the corresponding terms in the sensitivity matrix can also be evaluated analytically; for instance, if ${\bm a}_{\rm NG} = A_1 g(r) {\bm e}_r $,
\begin{align}
\frac{\partial {\bm a}_{\rm NG}}{\partial A_1} = \frac{{\bm a}_{\rm NG}}{A_1},
\end{align}
and similarly for the other cases.

\label{lastpage}
\end{document}